\newcommand{\TODO}[1]{\textcolor{red}{#1}\GenericWarning{}{LaTeX Warning: TODO: #1}}\newcommand\todo\TODO

\documentclass[sigconf,nonacm,screen,natbib=false]{acmart}
\acmConference[FSE Companion '24]{Companion Proceedings of the 32nd ACM International Conference on the Foundations of Software Engineering}{July 15--19, 2024}{Ipojuca (Pernambuco), Brazil}
\acmBooktitle{Companion Proceedings of the 32nd ACM International Conference on the Foundations of Software Engineering (FSE Companion '24), July 15--19, 2024, Ipojuca (Pernambuco), Brazil}

\usepackage{svg}
\usepackage{listings}
\usepackage{xcolor}

\usepackage{pifont}
\usepackage{multicol}
\usepackage{paralist}
\usepackage{diagbox}
\usepackage{multirow}
\newcommand{\cmark}{\ding{51}}%
\newcommand{\xmark}{\ding{55}}%

\usepackage[inline]{enumitem}

\definecolor{codegreen}{rgb}{0,0.6,0}
\lstdefinestyle{CoqStyle}{
    backgroundcolor=\color{white},   
    commentstyle=\color{codegreen},
    keywordstyle=\color{blue},
    numberstyle=\tiny\color{black},
    stringstyle=\color{red},
    basicstyle=\ttfamily\scriptsize,
    morekeywords={Require,Import,Proof,Qed,Lemma,Theorem,Admitted},
    breakatwhitespace=false,         
    breaklines=true,                 
    captionpos=b,                    
    keepspaces=true,                 
    numbers=left,                    
    xleftmargin=5.0ex,
    numbersep=5pt,                  
    showspaces=false,                
    showstringspaces=false,
    showtabs=false,                  
    tabsize=2
}

\AtBeginDocument{%
  }

\setcopyright{rightsretained}
\copyrightyear{2024}
\acmYear{2024}
\acmDOI{10.1145/3663529.3663814}
\acmISBN{979-8-4007-0658-5/24/07}

\RequirePackage[
  datamodel=acmdatamodel,
  style=acmnumeric,
  sortcites=true,
  ]{biblatex}

\makeatletter
\def\@copyrightspace{\relax}
\makeatother

\begin{document}

\title{CoqPyt: Proof Navigation in Python in the Era of LLMs}

\author{Pedro Carrott}
\authornote{Both authors contributed equally to this research.}
\email{pedro.carrott@imperial.ac.uk}
\orcid{0000-0003-4316-928X}
\affiliation{%
  \institution{Imperial College London}
  \city{London}
  \country{UK}
}

\author{Nuno Saavedra}
\authornotemark[1]
\email{nuno.saavedra@tecnico.ulisboa.pt}
\orcid{0000-0003-4148-5991}
\affiliation{%
  \institution{INESC-ID/IST, University of Lisbon}
  \city{Lisbon}
  \country{Portugal}
}

\author{Kyle Thompson}
\orcid{0000-0002-2868-7612}
\email{r7thompson@ucsd.edu}
\affiliation{%
  \institution{University of California San Diego}
  \city{San Diego}
  \country{USA}
}

\author{Sorin Lerner}
\orcid{0000-0003-3957-0628}
\email{lerner@cs.ucsd.edu}
\affiliation{%
  \institution{University of California San Diego}
  \city{San Diego}
  \country{USA}
}

\author{João F. Ferreira}
\orcid{0000-0002-6612-9013}
\email{joao@joaoff.com}
\affiliation{%
  \institution{INESC-ID/IST, University of Lisbon}
  \city{Lisbon}
  \country{Portugal}
}

\author{Emily First}
\orcid{0000-0002-2896-2928}
\email{emfirst@ucsd.edu}
\affiliation{%
  \institution{University of California San Diego}
  \city{San Diego}
  \country{USA}
}

\renewcommand{\shortauthors}{Carrott et al.}

\begin{abstract}


  Proof assistants enable users to develop machine-checked proofs regarding software-related properties.
  Unfortunately, the interactive nature of these proof assistants imposes most of the proof burden on the user, making formal verification a complex, and time-consuming endeavor.
  Recent automation techniques based on neural methods address this issue, but require good programmatic support for collecting data and interacting with proof assistants.
  This paper presents CoqPyt, a Python tool for interacting with the Coq proof assistant.
  CoqPyt improves on other Coq-related tools by providing novel features, such as the extraction of rich premise data.
  We expect our work to aid development of tools and techniques, especially LLM-based, designed for proof synthesis and repair.
  A video describing and demonstrating CoqPyt is available at: \url{https://youtu.be/fk74o0rePM8}.
\end{abstract}




\maketitle

\section{Introduction}


Formal software verification is an incredibly effective method for developing high quality software, as it ensures that a software program adheres to a predefined formal specification.
For example, a study compared industrial standard compilers (\textit{e.g.}, GCC and LLVM) to CompCert, a C compiler verified using the Coq proof assistant~\cite{coq}, and CompCert was the only one for which no bugs were found~\cite{Yang11a}.
Unfortunately, even though formal software verification provides valuable guarantees, its development is still too costly.
CompCert
took 100,000 lines of Coq code and 6 person-years to formally verify~\cite{leroy2016compcert}. 
For this reason, it is important to find methods that decrease the cost of formal verification. 

One recent approach to address this is the use of neural methods to automate proof synthesis. 
\emph{Neural theorem provers}, given a partial proof and the \emph{proof state}, use neural networks to predict the next likely \emph{proof steps}. A proof assistant is then used to evaluate the candidate proof steps, returning either new proof states or errors.
Neural theorem provers iterate on this procedure, performing \emph{proof search} to traverse the space of possible proofs.
A complementary task is automated proof repair~\cite{ringer2021proof}, which is necessary to perform when specifications or dependencies change and result in broken proofs.
To train these neural models and implement proof search, sufficient programmatic support is thus required for collecting training examples and driving the theorem prover.



In this paper, our target proof assistant is Coq due to its popularity for building verified software systems.
Existing tools, such as Coq Serapy~\cite{sanchez2020generating} and PyCoq~\cite{pycoq}, provide programmatic support for interacting with Coq in Python, and all current neural theorem provers for Coq utilize them. These tools implement Python bindings for Coq SerAPI~\cite{GallegoArias2016SerAPI}, which is a library for machine-to-machine interaction with Coq. However, with the announcement that Coq SerAPI will be deprecated and replaced with Coq LSP~\cite{coq-lsp}, a language server for Coq, the need arises for a new tool to continue supporting existing features. Fortunately, this also creates a great opportunity to support new features that prior tools did not.



In the emerging era of large language models (LLMs), additional functionality is desirable in tools for interfacing with Coq in Python in order to promote LLM-based approaches for Coq proof synthesis and repair. Namely, the ability to collect rich contextual data is not supported by any of the existing programmatic tools for controlling Coq.
LLM-based neural theorem provers for Lean and Isabelle proof assistants rely on \emph{retrieval augmentation}, wherein they retrieve premises, such as lemmas and definitions, that are relevant to the proof goal, and condition the next tactic generation on those premises~\cite{yang2023leandojo, mikuła2023magnushammer}. For the approach to be effective, they collect fine-grained annotations of which premises are used in proofs as well as which premises are accessible from a proof.
However, existing tools for Coq do not collect that type of premise data.

\begin{figure*}[t]
  \centering
  \includegraphics[width=\linewidth]{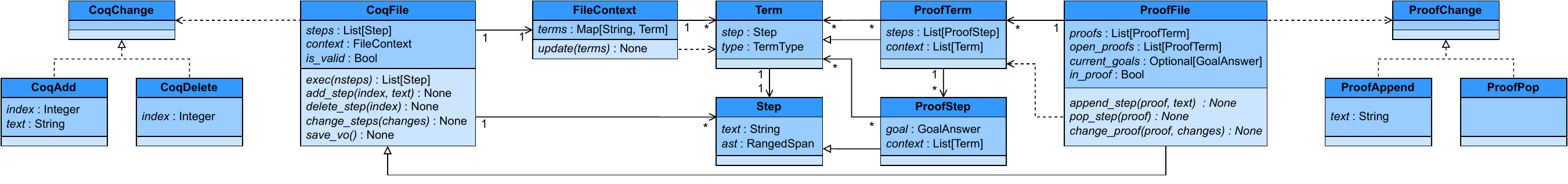}
  \caption{UML diagram for a simplified view of the domain of CoqPyt. The selected attributes and methods are all public. 
  }
  \label{fig:coqpyt}
\end{figure*}

We introduce CoqPyt, a tool that enables interaction with Coq in Python using Coq LSP as the backend.
CoqPyt continues to support existing capabilities essential to training data collection and proof search, but goes beyond state-of-the-art tools for Coq by supporting the ability to collect rich contextual data, among other novel features.
Our tool supports the newest versions of Coq (8.17 to 8.19), which will facilitate mining software repositories for new, enriched training and evaluation data. We foresee that our work will help promote future research into neural proof synthesis and repair for Coq, especially as LLM capabilities continue to evolve. CoqPyt is open-source and available at: \url{https://github.com/sr-lab/coqpyt}.



\section{CoqPyt}
\label{sec:coqpyt}

CoqPyt is a Python tool that enables developers and researchers to interact programmatically with Coq. It also allows users to extract textual and structural information from Coq files. We now describe the structure and main functionalities of CoqPyt, providing an illustrative example on how to interact with the tool.

\subsection{The Coq Proof Assistant}
Coq provides an interactive environment with a rich type system suitable for theorem proving. A theorem in Coq is a type definition, which can be proven by constructing a \emph{proof term} with the stated theorem type. Since writing a proof term directly is difficult, Coq allows users to write a \emph{proof script} consisting of a sequence of high-level \emph{tactics} (\textit{e.g.}, induction or reflexivity). Each tactic guides Coq in a search for the desired proof term, refining the state until no new obligations hold. After a tactic application, Coq displays the current \emph{proof state}, which includes the goals to prove and the local context. When executed, a complete proof script generates the proof term.

\subsection{Coq Language Server Protocol}

To interact with Coq files, we use Coq LSP \cite{coq-lsp}, a language server that provides Coq-related features. Coq LSP was developed to replace Coq SerAPI \cite{GallegoArias2016SerAPI}, a well-known tool in the Coq community that implements a protocol for interacting specifically with Coq files. Unlike Coq SerAPI, Coq LSP follows a standardized protocol for interacting with text documents known as Language Server Protocol (LSP) \cite{lsp}. LSP servers, such as Coq LSP, provide useful language features, such as auto complete. Client applications, such as IDEs, can use these features by running their own LSP client.

Besides the usual LSP features, Coq LSP provides Coq-specific functionality, namely requesting the proof state at any given point of a Coq file. Coq LSP also allows users to request the abstract syntax tree (AST) of the entire document, which corresponds to a list of \textit{steps} in the file, each with its own AST representation. A step can be a \textit{term definition} (\textit{e.g.}, definitions and theorems) or a \textit{proof step} (\textit{e.g.}, mid-proof tactic application), determined by the AST contents of each step. For erroneous Coq files, Coq LSP is able to ignore erroneous steps and evaluate the remaining steps of the file, allowing any subsequent non-erroneous steps to define new terms. Coq LSP also enables project-wise imports by saving compiled \texttt{.vo} files, which may also benefit from this error-ignoring approach, given that valid files may still import terms from invalid files.

\subsection{Interaction with Coq Files}
CoqPyt implements a client for Coq LSP providing the interface shown in Figure \ref{fig:coqpyt}. This client implementation is encapsulated within \texttt{CoqFile}, a class that abstracts the state of an actual Coq file, hiding the interaction with Coq LSP. For instance, the \texttt{is\_valid} boolean flag indicates if the underlying Coq file is valid and the \texttt{save\_vo} method allows the user to compile the file.

A \texttt{CoqFile} object provides an \texttt{exec} method which allows users to take an arbirary number of \textit{forward} and \textit{backward} steps through the file. The method returns a list of \texttt{Step}s corresponding to the steps that were executed or backtracked during the call. Each step contains its textual representation in the file and the AST contents.

Given that steps may be term definitions, a \texttt{FileContext} object is kept within the \texttt{CoqFile}, indexing by name all terms that have been defined until the last executed step. Any term defined inside a module has its name prefixed by the module path. Although notations are nameless terms, these are also stored in the context, indexed by the string pattern used to define them. For each defined term, a \texttt{Term} object is created containing its corresponding step, as well as its \textit{term type}. This type depends on whether the term is a theorem, a notation, a tactic or any other definable Coq construct.

\subsection{Proof Navigation}


With CoqPyt, it is also possible to track the proof state throughout the file. A \texttt{CoqFile} can be instantiated through the \texttt{ProofFile} subclass, which allows users to manage the file's proof context. As steps are taken, one can enter (or leave) proof mode, which will activate (or deactivate) the \texttt{in\_proof} boolean flag. If there are any on-going goals, these can be accessed via the \texttt{current\_goals} attribute. Goals are represented as a \texttt{GoalAnswer} object, which mimics the structure of goals in the Coq LSP response \cite{coq-lsp}.

In addition to allowing users to manage a file's proof context, \texttt{ProofFile} instances also fetch all of the file's \textit{imported} terms. Thus, a \texttt{ProofFile} captures terms in a file and in its dependencies. In practice, the \texttt{ProofFile} context is initialized by instantiating a \texttt{CoqFile} for each library imported by the file and extracting the context of each library to the new \texttt{ProofFile} instance. Since creating a \texttt{ProofFile} requires instantiating multiple \texttt{CoqFile}s, it is more expensive to create a \texttt{ProofFile} than a \texttt{CoqFile}.

A \texttt{ProofFile} holds information about all proofs found in the file until the last executed step. Proofs with open goals are kept in the \texttt{open\_proofs} attribute. Multiple open proofs occur in files with \textit{nested} proofs, as we need to open inner proofs before closing outer proofs. Each proof is a \texttt{ProofTerm}, a \texttt{Term} with a \texttt{context} attribute, listing all terms found in the current context used to define the proof term. This contextual information is valuable for neural-based models and is not accessible from prior Coq-related tools.

A \texttt{ProofTerm} also contains an ordered list of \texttt{ProofStep}s, corresponding to all steps taken until the proof is closed (or until the last executed step if the proof is open). The \texttt{ProofStep} class enables a mapping from each proof step to the intermediate proof goals it attempts to solve or simplify. A \texttt{ProofStep} is thus constructed by augmenting the respective \texttt{Step}
with the intermediate proof goals before that step is taken. Similarly to \texttt{ProofTerm}, a \texttt{ProofStep} also has a \texttt{context} attribute listing the terms used in the step.

\subsection{Proof Modification}

CoqPyt allows steps to be \textit{added to} or \textit{deleted from} the file. Allowing such modifications on Coq files yields an interface well suited for proof development. Invalid changes, such as adding inexisting tactics, may lead to explicit errors or undefined behaviour, depending on the nature of the error. For erroneous modifications, the file will revert to its original state, ignoring the requested changes.

Through the \texttt{add\_step} method, it is possible to define terms, introduce theorems and apply tactics to solve proof obligations. Conversely, \texttt{delete\_step} enables the removal of steps, such as added tactics which do not alter the proof goals as intended. Both methods allow editing in arbitrary parts of the file: \texttt{delete\_step} requires an index \texttt{i} to delete the \texttt{i}$^\text{th}$ step in the file; \texttt{add\_step} requires the index of the step which precedes the new step. To add a new step, its textual representation must also be provided.

Modifications may also be batched as transactions. A sequence of \texttt{CoqChange}s can be provided to the \texttt{change\_steps} method, which will perform all modifications. A \texttt{CoqChange} can be either a \texttt{CoqAdd} or a \texttt{CoqDelete}, which serve as data wrappers for the parameters of \texttt{add\_step} and \texttt{delete\_step}, respectively. During the transaction, intermediate invalid states are allowed as long as the final state is valid. For example, it is possible to add a \texttt{Qed} before adding the actual proof steps, which cannot be done through successive \texttt{add\_step} calls. To directly modify a proof, \texttt{ProofFile} provides the methods \texttt{append\_step} and \texttt{pop\_step} for single step modifications, as well as a \texttt{change\_proof} method for a transaction of \texttt{ProofChange}s. 

\subsection{Use Cases}
The features that CoqPyt offers are particularly useful for proof search and learning-based tasks, such as neural theorem proving.
To get a sense of CoqPyt's interface for collecting data and conducting proof search, consider the file \texttt{test.v} in Listing~\ref{lst:coq-file}, where we have a Coq file that defines a property about reversing a list.
Listing~\ref{lst:py-data} shows how we can leverage CoqPyt to collect data from the file shown in Listing~\ref{lst:coq-file}. 
We can see which terms are available to the file via the \texttt{pf.context} attribute (line 3).
For example, all the terms from the imported package \texttt{List.v} are available to the proofs in the file.
Likewise, we show how to retrieve information from each proof step in the file via the \texttt{ProofStep} attributes (line 6). 

\lstset{style=CoqStyle}
\begin{lstlisting}[caption={Example Coq file \texttt{test.v} with a property of \texttt{rev}.}, label={lst:coq-file}, float]
Require Import List.
Lemma rev_append: forall {a} (l1 l2: list a),
  rev (l1 ++ l2) = rev l2 ++ rev l1.
Proof.
intros a l1 l2. induction l1; intros. 
  - simpl. rewrite app_nil_r. reflexivity.
  - simpl. rewrite IHl1.
Admitted.
\end{lstlisting}

\begin{lstlisting}[caption={Proof data available with CoqPyt.},label={lst:py-data},language=Python,float]
with ProofFile("test.v") as pf:
  pf.exec(len(pf.steps))
  print(pf.context)
  for proof in pf.proofs:
    for step in proof.steps:
      print(step.text, step.ast, step.context, step.goals)
\end{lstlisting}

Developing neural-based theorem provers requires the ability to attempt possibly erroneous proofs during proof search.
For example, suppose we wanted to complete the proof \texttt{rev\_append} from Listing~\ref{lst:coq-file}.
Listing~\ref{lst:attempts} shows how we can use the \texttt{change\_proof} method to conduct proof attempts.
We first obtain the \texttt{ProofTerm} object of the proof \texttt{rev\_append} (line 5). We then delete the \texttt{``Admitted.''} step from the proof of \texttt{rev\_append} (line 7) and add the steps associated with our new proof attempt (line 9).
When we attempt to add the erroneous steps from the proof attempt \texttt{incorrect} (line 11), an error is returned and \texttt{test.v} will remain in its original state.
In turn, when the valid steps from the proof attempt \texttt{correct} are added (line 11), the changes are applied to \texttt{test.v}.

\begin{lstlisting}[caption={Proof attempts with CoqPyt.},label={lst:attempts},language=Python,float]
incorrect = [" reflexivity.", "\nQed."]
correct = [" rewrite app_assoc."] + incorrect
with ProofFile("test.v") as pf:
    pf.exec(len(pf.steps))
    unproven = pf.unproven_proofs[0]
    for attempt in [incorrect, correct]:
        changes = [ProofPop()]  # Admitted
        for s in attempt:
            changes.append(ProofAppend(s))
        try:
            pf.change_proof(unproven, changes)
            print("Proof succeeded!")
            break
        except InvalidChangeException:
            print("Proof attempt not valid.")
\end{lstlisting}

\begin{table*}[t]
  \centering
  \caption{Evaluation of CoqPyt's performance on CompCert. PCC stands for Pearson correlation coefficient.}
  \label{tab:performance_eval}
  \footnotesize
  \begin{tabular}{lcccccccccccc}
    \toprule
    & \multicolumn{2}{c}{Count} & \multicolumn{10}{c}{Execution time (s)}\\
    \cmidrule(lr){2-3} \cmidrule(lr){4-13}
    \multirow{2}{*}{Metric/Feature} & & & & & \multicolumn{2}{c}{\textbf{Add step}} & \multicolumn{2}{c}{\textbf{Delete step}} & \multicolumn{2}{c}{\textbf{Change steps (Add)}} & \multicolumn{2}{c}{\textbf{Change steps (Delete)}} \\
    \cmidrule(lr){6-7} \cmidrule(lr){8-9} \cmidrule(lr){10-11} \cmidrule(lr){12-13}
    & \textbf{Proofs} & \textbf{Steps} & \textbf{Load file} & \textbf{Execute file} & \tiny{Beginning} & \tiny{On pointer} & \tiny{Beginning} & \tiny{On pointer} & \tiny{Beginning} & \tiny{On pointer} & \tiny{Beginning} & \tiny{On pointer} \\
    \midrule
    $\mu (\sigma^2)$ & 32.3 (48.1) & 609.6 (828.1) & 6.6 (7.3) & 31.2 (63.8) & 1.1 (1.8) & 0.9 (1.5) & 1.0 (1.7) & 0.9 (1.6) & 1.5 (2.7) & 1.0 (1.7) & 1.1 (1.8) & 0.8 (1.5) \\
    $\tilde{x}$              & 12.0 & 258.0 & 3.9 & 9.6 & 0.4 & 0.3 & 0.3 & 0.3 & 0.5 & 0.3 & 0.4 & 0.3 \\
    \midrule
    PCC (steps) & -     & 1.00   & 0.59 & 0.72 & 0.90 & 0.92 & 0.94 & 0.92 & 0.87 & 0.92 & 0.90 & 0.91 \\
    \bottomrule
  \end{tabular}
\end{table*}

\section{Related Work}
\label{sec:relatedwork}

Most neural theorem provers for Coq use either Coq Serapy~\cite{sanchez2020generating, thakur2023languageagent}, PyCoq~\cite{tactician}, or a custom Python class~\cite{yang2019learning, First20oopsla, Sanchez-Stern22passport, First22icse} to create training examples and perform proof search. The neural theorem prover GamePad~\cite{Huang2019Gamepad} instead modifies Coq itself to record intermediate proof states. The CoqGym benchmark dataset~\cite{yang2019learning}, collected using a Python class, is the state-of-the-art benchmark for Coq. However, it was collected in 2019 and has only been updated to support versions 8.10 and 8.12 of Coq, while the newest version is 8.19. CoqPyt will render neural theorem provers useful for new Coq developments since it can be used for mining software repositories for newer Coq training data and for evaluation. 



LeanDojo~\cite{yang2023leandojo} and PISA~\cite{jiang2021lisa} are learning environments for Lean and Isabelle, allowing for data extraction and proof search interaction. The neural theorem provers that use these environments are LLM-based and use retrieval augmentation~\cite{Jiang2022Thor, mikuła2023magnushammer,yang2023leandojo}. CoqPyt would enable retrieval augmented LLM approaches in Coq. 



Pretrained LLMs have been shown to have quantitative reasoning capabilities~\cite{palm, touvron2023llama, openai2023gpt4}, especially when they undergo continued training on math data~\cite{Lewkowycz2022Minerva,azerbayev2023llemma}. To synthesize proofs, they can be fine-tuned on proof data~\cite{first2023baldur, yang2023leandojo, jiang2021lisa}, few-shot prompted~\cite{azerbayev2023llemma, jiang2023draft, zhang2023getting}, or even zero-shot prompted~\cite{thakur2023languageagent,yang2023leandojo}.
However, the question of test leakage arises in any evaluation that uses pretrained LLMs. CoqPyt will allow for data collection after the pretraining cutoff date.


During proof development, proof engineers are constantly performing proof repair~\cite{ringer2020replica}.
The first work in automating this task uses symbolic tools for automated proof repair in Coq~\cite{ringer2021proof},
and has since been applied to other proof systems~\cite{masci2022proof}. Baldur is the first work to use neural methods (in particular, LLMs) to repair proofs, but as part of its proof synthesis approach~\cite{first2023baldur}. With the creation of a large-scale dataset of proof repair instances~\cite{reichel2023proof}, the creation of LLM methods to automate real-world proof repairs is forthcoming.

\section{Evaluation}
\label{sec:evaluation}


\begin{table}[t]
  \centering
  \caption{Feature comparison between CoqPyt and similar tools (\cmark full support, $\sim$ partial support, \xmark\ no support). 
  }
  \label{tab:feature_comparison}
  \footnotesize
  \begin{tabular}{lccc}
    \toprule
    \textbf{Feature} & \textbf{CoqPyt} & \textbf{Coq LSP} & \textbf{Coq Serapy}\\
    \midrule
    Get proof state & \cmark & \cmark & \cmark \\
    Check file validity & \cmark & \cmark & \cmark \\
    Execute/Modify steps & \cmark/\cmark & $\sim$/\xmark & \cmark/$\sim$ \\
    Extract step context & \cmark & \xmark & \xmark \\
    Track modules/terms/proofs & \cmark/\cmark/\cmark & \xmark/\cmark/\xmark & \cmark/\xmark/$\sim$ \\
    \bottomrule
  \end{tabular}
\end{table}

\paragraph{Feature Evaluation} Table \ref{tab:feature_comparison} compares the features of CoqPyt to similar tools. When compared to Coq LSP, CoqPyt supports a more extensive range of features by using the data supplied by Coq LSP and constructing additional components of Coq program logic upon it. For instance, Coq LSP provides partial support for operating with steps by enabling retrieval of the proof state at any position of the file. However, it lacks implementation for navigating backward or forward from a specific step, which is provided by CoqPyt. Compared to Coq Serapy, CoqPyt offers complete support for an additional four features:
\begin{inparaenum}
    \item modifying steps -- while Coq Serapy supports the addition or deletion of steps at the current point of execution, it does not offer the capability to add or delete steps at arbitrary positions within the file;
    \item extract step context -- Coq Serapy does not collect the terms used in each step;
    \item track terms -- Coq Serapy does not retain a record of the terms defined up to the current execution point;
    \item track proofs -- even though Coq Serapy supports commands such as navigating to the next proof, it does not maintain a record of all the proofs already defined in a file.
\end{inparaenum}

\paragraph{Performance Evaluation} To evaluate CoqPyt's performance, we used the project CompCert~\cite{leroy2016compcert} since it is a popular project with over 1,700 stars on GitHub and 259 Coq files with considerable complexity. We ran the experiments inside a Docker container on an Intel Xeon Silver 4210R CPU @ 2.40GHz. Table \ref{tab:performance_eval} summarizes our results. CoqPyt was able to run on 193 files, 58 files contained Coq-related errors, 6 files generated out-of-memory errors, and 2 files crashed for unknown reasons. The column \emph{Count} describes metrics for the number of steps and proofs of CompCert. We ran all the features in each Coq file. For features that modify steps, we performed the modification in the beginning of the file and at the current point of execution, after executing the whole file. We performed each case 5 times for each feature. This approach is driven by the understanding that the number of required operations for these features increases as the distance to the point of execution rises. To evaluate $\texttt{change\_steps}$, we performed a single addition or deletion to make it simpler to compare to the other features. The results for file execution are dependent on a cache that saves our loading of Coq libraries. As more libraries were cached, the time to execute the files decreased. By calculating the Pearson correlation coefficient, we conclude that the execution time for all features is positively correlated to the number of steps in a file. A replication package is available at: \url{https://zenodo.org/records/10292580}.
 


\section{Conclusion}
\label{sec:conclusion}
We developed CoqPyt, a Python tool for interacting with Coq. It offers valuable features previously unavailable in similar Coq-related tools, namely per-step context extraction and arbitrary file and proof modification. CoqPyt does not yet explore the full potential of context extraction, as it does not consider a small subset of Coq's rich environment (\textit{e.g.}, module types and section-local constructs) nor does it fetch context recursively, \textit{i.e.}, for each term in a step context we ignore the context of the term itself. Nonetheless, we have shown that CoqPyt provides rich features and reasonable execution times, enabling an expressive interaction style with Coq. These contributions are expected to ease the automation of machine-checked proof synthesis and repair in the emerging era of LLMs.







\section*{Acknowledgements}
This work was partially funded by NSF, National Science Foundation grants CCF-1955457 and CCF-2220892, and by FCT, Fundação para a Ciência e a Tecnologia, under grant BD/04736/2023 and project UIDB/50021/2020 (DOI:10.54499/UIDB/50021/2020).


\balance
\printbibliography

\end{document}